\documentclass[preprint]{article}
\usepackage{authblk}
\usepackage{graphicx}% Include Figure files
\usepackage{dcolumn}% Align table columns on decimal point
\usepackage{bm}% bold math
\usepackage{amsmath}
\usepackage{amssymb}
\usepackage{multirow}
\usepackage{caption}
\usepackage{subcaption}
\usepackage{epstopdf}
\usepackage{hyperref}
\usepackage{cite}
\begin{document}
\title{\bf Cosmological insights from an exponential $Om(z)$ function in $f(T,T_{G})$ gravity framework}
\author[]{Amit Samaddar\thanks{samaddaramit4@gmail.com}}
\author[]{S. Surendra Singh\thanks{ssuren.mu@gmail.com}}
\affil[]{Department of Mathematics, National Institute of Technology Manipur, Imphal-795004,India.}

\maketitle
 
 \textbf{Abstract}:We examine a modified teleparallel gravity model defined by $f(T,T_{G})=T+\gamma\sqrt{T_{G}}+\delta\sqrt{T}$ by introducing an exponential $Om(z)$ diagnostic of the form $Om(z)=\alpha e^{\frac{z}{1+z}}+\beta$. This novel form captures smooth redshift evolution and allows for a flexible, model-independent probe of dark energy dynamics. We derive a Hubble function from this expression and use MCMC analysis with $31$ CC, $26$ BAO and $1701$ Pantheon+ data points to constrain the model parameters. The best-fit results yield $H_{0} \in  [68.46, 77.38]$ km/s/Mpc for  $\alpha  \in [-0.232, -0.068]$ and $\beta \in [0.218, 0.560]$ which is consistent with local $H_{0}$ values. Our model predicts a transition redshift $z_{tr} \approx( 0.48- 0.54)$, present $q_{0}\approx - 0.34$, and $\omega_{0}\approx - 0.33$. It satisfies NEC and DEC, closely tracks $\Lambda$CDM in the statefinder plane and estimates a cosmic age of $(13.28-13.87)$ Gyr which confirms its strength in explaining late-time acceleration. Our findings demonstrate that the exponential $Om(z)$ parameterization provides a robust and insightful approach to trace dark energy evolution within modified gravity frameworks.
 
 \textbf{Keywords}: Modified gravity, $f(T,T_{G})$ gravity, exponential $Om(z)$ form, observational cosmology, statefinder analysis, late-time Universe.
 
 \section{Introduction}\label{sec1}
 \hspace{0.5cm} The discovery that our Universe is undergoing a phase of accelerated expansion stands as one of the most significant breakthroughs in modern cosmology. This observation was first inferred from luminosity-distance measurements of Type Ia supernovae \cite{Riess98,Per99}, and has since been corroborated by several independent observational probes, including measurements of the cosmic microwave background (CMB) anisotropies \cite{WMAP13}, baryon acoustic oscillations (BAO) \cite{Dan05} and large-scale structure surveys \cite{Tegmark04}. Within the standard cosmological model, based on general relativity (GR), this acceleration is typically ascribed to an exotic component known as dark energy (DE), characterized by a strong negative pressure. The most straightforward interpretation of dark energy is a cosmological constant $\Lambda$, representing vacuum energy with a constant equation of state $\omega=-1$. While the $\Lambda$CDM model remains remarkably consistent with most observational data, it faces substantial theoretical challenges—most notably, the cosmological constant problem, which refers to the vast discrepancy between the observed value of $\Lambda$ and theoretical predictions from quantum field theory \cite{Steven89}, and the coincidence problem, which questions why the densities of dark energy and matter are of the same order precisely in the current epoch \cite{PJ99}. These open issues suggest that a deeper understanding of gravity, possibly through geometric modifications to GR, may be necessary to explain the origin of cosmic acceleration.
 
 The discovery of the late-time acceleration of the Universe has motivated the development of modified theories of gravity that go beyond the standard Einstein–Hilbert action. These theories aim to explain cosmic acceleration without invoking exotic dark energy components. In the curvature-based approach, generalizations such as $f(R)$ gravity \cite{Nojiri11,ADEF10}, $f(G)$ gravity involving the Gauss-Bonnet invariant \cite{Nojiri05,Bamba11} and more general $f(R,G)$ models \cite{TPS10}, have been widely studied. These retain the Levi-Civita connection and rely on curvature as the geometric quantity describing gravity. Alternatively, a compelling and fundamentally different approach involves extending the torsion-based formulation of gravity. It has been known since Einstein’s time that one can construct the Teleparallel Equivalent of General Relativity (TEGR) by replacing the torsionless Levi-Civita connection with the curvature-free Weitzenb$\ddot{o}$ck connection \cite{Hayashi79,Maluf94,Ald13,Maluf13}. In this formulation, gravity is described by the torsion scalar $T$, which plays a role analogous to the Ricci scalar $R$ in General Relativity (GR). This allows for the construction of $f(T)$ gravity theories \cite{Ferr07,Ferr08,GRB09}, which although TEGR and GR are equivalent, differ fundamentally from $f(R)$ theories due to the nature of the connection and torsion formulation.

Building on this torsional framework, more complex models have been emerged which incorporate higher-order torsion contributions, analogous to the role of $G$ in curvature-based gravity. One notable advancement is the construction of the teleparallel equivalent of the Gauss–Bonnet term, denoted $T_{G}$ a quartic scalar invariant that reduces to a topological term in four dimensions \cite{Boul85,Wheel86}. This development led to the formulation of $f(T,T_{G})$ gravity \cite{Ant94,Kanti99,SNO05,Kadam23,Sharif18}, a novel and rich class of theories that extends $f(T)$ gravity by including higher-order torsional invariants.

To evaluate the viability of alternative cosmological models, it is crucial to employ diagnostics that are independent of specific assumptions about dark energy. One such diagnostic is the $Om(z)$ function, which has emerged as a robust and model-independent tool for differentiating the standard $\Lambda$CDM cosmology from models with evolving dark energy. This diagnostic is constructed using the Hubble parameter $H(z)$ and provides insights into the dynamics of the Universe’s expansion history. For the $\Lambda$CDM model, $Om(z)$ remains constant across redshift, whereas deviations from constancy signal the presence of dynamical dark energy effects \cite{Sahni2008,Shafieloo2012}. Importantly, the redshift evolution of $Om(z)$ allows one to infer the nature of the dark energy equation of state $\omega(z)$: an increasing $Om(z)$ with redshift is typically associated with phantom behavior ($\omega<-1$), while a decreasing trend indicates quintessence-like characteristics ($\omega>-1$) \cite{Zunckel2008}. Numerous studies in the literature have proposed different parameterizations for the $Om(z)$ function to better capture the dynamics of cosmic acceleration \cite{Sahni14,Ding15,XZHeng16,Qi18,Myrz23,YMyr25}. As such, $Om(z)$ serves as a valuable complementary probe to direct measurements of $H(z)$, enabling a consistency check on the cosmological constant hypothesis and providing constraints on more general dark energy scenarios. In this work, we propose a new exponential parameterization given by: $Om(z)=\alpha e^{\frac{z}{1+z}}+\beta$, which is smooth and well-behaved across redshifts. This functional form allows for a gradual transition from early-time deceleration to late-time acceleration and provides sufficient flexibility to fit current observational datasets. We reconstruct the Hubble parameter $H(z)$ based on this $Om(z)$ form and constrain the parameters $\alpha$, $\beta$ and $H_0$ using a comprehensive cosmological dataset. A Markov Chain Monte Carlo (MCMC) analysis is carried out to determine the best-fit values and uncertainties. Our results indicate that the exponential $Om(z)$ parameterization is in excellent agreement with observations. It describes a consistent evolution of the deceleration parameter $q(z)$ and the effective equation of state $\omega(z)$, satisfies the classical energy conditions, and estimates a value of $H_0$ compatible with both local and CMB measurements. Moreover, the statefinder diagnostic shows that the model evolution closely follows that of $\Lambda$CDM, reinforcing its credibility within the context of $f(T,T_{G})$ gravity as a promising alternative framework to explain late-time cosmic acceleration.

The structure of this paper is as follows: In section \ref{sec2}, we present the field equations in the framework of $f(T,T_G)$ gravity. Section \ref{sec3} is devoted to the discussion of a specific $f(T,T_G)$ model and its cosmological implications. In section \ref{sec4}, we introduce the $Om(z)$ diagnostic and outline the motivation behind adopting its exponential parameterization. Section \ref{sec5} details the observational datasets and the MCMC analysis used to constrain the model parameters. In section \ref{sec6}, we examine key cosmological parameters derived from the best-fit results. Section \ref{sec7} is dedicated to the investigation of energy conditions and section \ref{sec8} presents a statefinder diagnostic to further characterize the model. In section \ref{sec9}, we calculate the age of the Universe predicted by the model. Finally, our conclusions are summarized in section \ref{sec10}.
 \section{Formulation of $f(T,T_{G})$ gravity and derivation of field equations}\label{sec2}
 \hspace{0.5cm} Teleparallel Gravity (TG) employs a connection that is curvature-free, differing fundamentally from GR, where the Levi-Civita connection induces spacetime curvature. In the TG framework, the role of the Levi-Civita connection $\overset{\circ}\Gamma\;^{\theta}_{\mu\nu}$ is replaced by the Weitzenböck connection $\Gamma^{\theta}_{\mu\nu}$, which is constructed to have vanishing curvature but non-zero torsion \cite{Hayashi79}. The core dynamical entities in TG are the tetrad fields $e^{\varrho}_{\mu}$, along with their inverses $E^{\mu}_{\varrho}$, rather than the metric tensor alone. These tetrads serve as a bridge between the spacetime manifold and the tangent space and the metric tensor $g_{\mu\nu}$ is reconstructed from them via the relation \cite{Baha23}
 \begin{equation}\label{1}
 g_{\mu\nu}=e^{\varrho}_{\mu}e^{\varsigma}_{\nu}\eta_{\varrho\varsigma},
 \end{equation}
 where $\eta_{\varrho\varsigma}$ is the Minkowski metric defined as: $\eta_{\varrho\varsigma}=E^{\mu}_{\varrho}E^{\nu}_{\varsigma}$. In this formalism, Latin indices are used to represent components in the local Lorentz (or tangent) space, while Greek indices correspond to spacetime coordinates \cite{Cai16}. Although tetrads can be utilized in curvature-based gravity theories like GR, their explicit role is often downplayed, likely because the associated connections are not torsion-free in those settings. In contrast, teleparallel gravity employs tetrads as fundamental fields that define a globally flat, torsion-rich geometry. A critical property of the tetrad fields is their mutual orthogonality, captured by the relations
 \begin{equation}\label{2}
 e^{\varrho}_{\mu}E^{\mu}_{\varsigma}=\delta^{\varrho}_{\varsigma}, \hspace{0.4cm} e^{\varrho}_{\mu}E^{\nu}_{\varrho}=\delta^{\nu}_{\mu},
 \end{equation}
The Weitzenb$\ddot{o}$ck connection, characteristic of teleparallel gravity, combines tetrad fields and spin connections in its construction. It is given by the expression
\begin{equation}\label{3}
\Gamma^{\theta}_{\mu\nu}=E^{\theta}_{\varrho}\big[\partial_{\mu}e^{\varrho}_{\nu}+w^{\varrho}_{\varsigma\mu} e^{\varsigma}_{\nu}\big],
\end{equation}
Here, the tetrad encapsulates the metric structure of spacetime and the spin connection $w^{\varrho}_{\varsigma\mu}$ accounts for the six independent components associated with local Lorentz transformations.  Together, these variables describe both the gravitational dynamics and local inertial effects in a covariant way. Since the teleparallel connection is constructed to be curvature-free, the standard Riemann tensor identically vanishes in this formalism. As a result, the geometric information of gravity is instead captured by the torsion tensor, which plays a central role in describing gravitational interactions in this approach. The torsion tensor $T^{\theta}_{\mu\nu}$ quantifies the antisymmetric part of the connection, and is formally defined as \cite{RWE23}
\begin{equation}\label{4}
T^{\theta}_{\mu\nu}=2\Gamma^{\theta}\;_{[\nu\mu]},
\end{equation}
where square brackets denote antisymmetrization of indices. This tensor is fundamental in teleparallel gravity, as it replaces curvature as the descriptor of gravitational interaction. To bridge the gap between the teleparallel and curvature-based formalisms, one introduces the contortion tensor $K^{\theta}_{\mu\nu}$, defined by the difference between the Weitzenb$\ddot{o}$ck and Levi-Civita connections \cite{Clifton12}:
\begin{equation}\label{5}
K^{\theta}_{\mu\nu}=\Gamma^{\theta}_{\mu\nu}-\overset{\circ}\Gamma\;^{\theta}_{\mu\nu}=\frac{1}{2}\big[T_{\mu\;\;\nu}^{\;\;\theta}+T_{\nu\;\;\mu}^{\;\;\theta}-T_{\;\;\mu\nu}^{\theta}\big],
\end{equation}
The contortion tensor encodes how torsion modifies the geometry relative to standard GR, and it is crucial in constructing scalar invariants. From the torsion tensor and its contractions, one can construct the torsion scalar $T$, which serves as the Lagrangian density in the teleparallel equivalent of GR and its extensions \cite{Krssak19}.
\begin{equation}\label{6}
T=\frac{1}{4}T_{\;\;\mu\nu}^{\sigma}T_{\sigma}^{\;\;\mu\nu}+\frac{1}{2}T_{\;\;\mu\nu}^{\sigma}T_{\;\;\;\;\sigma}^{\nu\mu}-T_{\;\;\mu\sigma}^{\sigma}T_{\;\;\;\;\psi}^{\psi\mu}
\end{equation}
Teleparallel gravity's action is built from the torsion scalar $T$, yielding dynamical equations equivalent to those from GR's Einstein-Hilbert action. In the teleparallel framework, the Ricci scalar $R$ derived from the connection is identically zero, $R=0$. However, one can still express the standard Ricci scalar $\overset{\circ}R$, derived using the Levi-Civita connection $\overset{\circ}\Gamma\;^{\theta}_{\mu\nu}$ in terms of torsional quantities via the relation \cite{Baha15,Amit23}:
\begin{equation}\label{7}
\overset{\circ}R=B-T,
\end{equation}
where $B$ is a boundary term that contributes only as a total divergence. This boundary term is explicitly given by:
\begin{equation}\label{8}
B=\frac{2}{e}\partial_{\alpha}\bigg[eT^{\mu\;\;\alpha}_{\;\;\;\mu}\bigg],
\end{equation}
Here, $e=| e^{A}_{\mu}|=\sqrt{-g}$ denotes the determinant of the tetrad matrix. The relation established in equation (\ref{7}) highlights that the field equations derived from GR and its teleparallel equivalent TEGR are identical at the classical level, confirming their dynamic equivalence. Beyond the torsion scalar, another significant curvature-based invariant is the Gauss-Bonnet term $G$, expressed as \cite{Kof14,SCap16}:
\begin{equation}\label{9}
G=\overset{\circ}{R^{2}}-4\overset{\circ}R_{\mu\nu}\overset{\circ}{R}{}^{\mu\nu}+\overset{\circ}R_{\mu\nu\sigma\psi}\overset{\circ}{R}{}^{\mu\nu\sigma\psi},
\end{equation}
which combines Ricci scalar, Ricci tensor and Riemann tensor contractions. In four dimensions, this term does not contribute dynamically in GR due to its topological nature, but in modified gravity scenarios or higher-dimensional theories, it plays a vital role. In the framework of teleparallel gravity, a torsional counterpart to the Gauss-Bonnet term can be formulated, denoted by $T_{G}$.
\begin{eqnarray}\label{10}
T_{G}&=&\bigg[K_{A\;\;e}^{\;\;\;\beta}K_{B}^{\;\;e\gamma}K_{C\;\;\;F}^{\;\;\;\varphi}K_{D}^{\;\;F\lambda}-2K_{A}^{\;\;\beta\gamma}K_{B\;\;e}^{\;\;\;\varphi}K_{C\;\;F}^{\;\;\;e}K_{D}^{\;\;F\lambda}
+2K_{A}^{\;\;\beta\gamma}K_{B\;\;e}^{\;\;\;\varphi}K_{F}^{\;\;e\lambda}K_{D\;\;C}^{\;\;\;F}\\\nonumber
&&+2K_{A}^{\;\;\beta\gamma}K_{B\;\;e}^{\;\;\;\varphi}K_{C,D}^{\;\;\;e\lambda}\bigg]\delta^{ABCD}_{\beta\varphi\lambda\varphi},
\end{eqnarray}
In this formulation, the symbol $\delta^{ABCD}_{\beta\varphi\lambda\varphi}$ stands for the generalized Kronecker delta, which plays a key role in contracting antisymmetric indices in expressions involving the torsion-based Gauss-Bonnet term. This object can be explicitly expressed using the Levi-Civita tensor as
\begin{equation}\label{11}
\delta^{ABCD}_{\beta\varphi\lambda\varphi}=\epsilon^{ABCD}\epsilon_{\beta\gamma\varphi\lambda},
\end{equation}
This term reproduces the standard Gauss-Bonnet invariant modulo a total derivative, ensuring dynamical equivalence in the action.
\begin{equation}\label{12}
eB_{G}=\delta^{ABCD}_{\beta\gamma\varphi\lambda}\partial_{A}\bigg\{K_{B}^{\;\;\beta\gamma}\bigg[K_{C\;\;\;,D}^{\;\;\varphi\lambda}+K_{D\;\;C}^{\;\;\;I}K_{I}^{\;\;\varphi\lambda}\bigg]\bigg\},
\end{equation}
By uniting the torsion scalar with the boundary correction, one obtains the teleparallel equivalent of the Gauss-Bonnet invariant.
\begin{equation}\label{13}
G=B_{G}-T_{G},
\end{equation}
In analogy with the modifications made to GR, we extend the Einstein-Hilbert action by incorporating general functions of both the torsion scalar and the teleparallel Gauss-Bonnet term. Given its relevance to early-Universe inflation and late-time cosmic acceleration, the Gauss-Bonnet contribution is a natural candidate for such a generalization. Assuming $\kappa^{2}=8\pi G$ and incorporating the matter sector via the Lagrangian $\mathcal{L}_{m}$, the action is expressed as follows \cite{Hoh17,Gkof14}:
\begin{equation}\label{14}
S=\frac{1}{2\kappa^{2}}\int d^{4}x\; e f(T,T_{G})+\int d^{4}x\; e\mathcal{L}_{m},
\end{equation}
In the limiting case where $f(T,T_{G})\rightarrow -T$, the framework simplifies to the TEGR. Within this modified approach, the torsion scalar contributes second-order terms, while the Gauss–Bonnet invariant introduces fourth-order corrections into the gravitational field equations. When considering a cosmological setting that is spatially flat, homogeneous, and isotropic, an appropriate choice for the tetrad field is given by \cite{LKD23}:
\begin{equation}\label{15}
e^{\varrho}_{\mu}=diag(1,a(t),a(t),a(t)),
\end{equation}
Here, $a(t)$ denotes the scale factor characterizing a spatially flat, homogeneous and isotropic Universe. This leads to the FLRW metric expressed as:
\begin{equation}\label{16}
ds^{2}=-dt^{2}+a(t)^{2}(dx^{2}+dy^{2}+dz^{2}),
\end{equation}
which aligns with the tetrad choice presented in equation (\ref{1}). The specific tetrad form corresponds to the Weitzenb$\ddot{o}$ck gauge, under which the spin connection vanishes, i.e., $w^{\varrho}_{\varsigma\mu}=0$. The torsion scalar and teleparallel Gauss-Bonnet invariant can be simplified using the earlier equations (see equations (\ref{6})–(\ref{10})) to the following forms:
\begin{equation}\label{17}
T=6H^{2}, \hspace{0.5cm} T_{G}=24H^{2}(\dot{H}+H^{2}),
\end{equation}
This shows that the teleparallel Gauss-Bonnet term and its curvature-based counterpart are equivalent in this cosmological context. Building upon the background dynamics discussed above, one can derive the modified Friedmann equations from the variation of the action with respect to the tetrad components. These equations incorporate the influence of both the torsion scalar $T$ and the teleparallel equivalent of the Gauss-Bonnet term $T_{G}$. The resulting field equations in a flat FLRW Universe take the form:
\begin{equation}\label{18}
\kappa^{2}\rho=\frac{f}{2}-6H^{2}f_{T}-\frac{T_{G}f_{T_{G}}}{2}+12H^{3}\dot{f}_{T_{G}},
\end{equation}
\begin{equation}\label{19}
\kappa^{2}p=-\frac{f}{2}+2(\dot{H}+3H^{2})f_{T}+2H\dot{f}_{T}+\frac{T_{G}f_{T_{G}}}{2}-\frac{T_{G}\dot{f}_{T_{G}}}{3H}-4H^{2}\ddot{f}_{T_{G}}.
\end{equation}
Here, a dot denotes differentiation with respect to the cosmic time $t$ and $f_{T}=\frac{\partial f}{\partial T}$, $f_{T_{G}}=\frac{\partial f}{\partial T_{G}}$. These equations govern the evolution of the Hubble parameter under the influence of generalized torsional gravity, incorporating both second- and fourth-order contributions via $T$ and $T_{G}$, respectively.
For simplicity and without loss of generality in our cosmological analysis, we adopt natural units by setting $\kappa^{2}=1$ in the subsequent calculations.
\section{Proposed $f(T,T_{G})$ model and its theoretical implications}\label{sec3}
\hspace{0.5cm} To examine the effects of torsional modifications to gravity on cosmic evolution, we choose a specific model for the function $f(T,T_{G})$, expressed as:
\begin{equation}\label{20}
f(T,T_{G})=T+\gamma\sqrt{T_{G}}+\delta\sqrt{T},
\end{equation}
where $\gamma$ and $\delta$ are model parameters. This model extends the TEGR, which is recovered in the limit $\gamma=\delta=0$, ensuring consistency with GR in appropriate regimes. The $\sqrt{T_{G}}$ term brings in higher-order torsional modifications akin to the Gauss-Bonnet invariant, which could account for early and late-time Universe phenomena without requiring a cosmological constant. The modification of the torsion scalar by the $\sqrt{T}$ term expands the Universe's evolutionary possibilities, yielding more varied phenomenology. This structure is not unique to $f(T,T_{G})$ theories; similar forms have been explored in a variety of extended theories of gravity. In $f(T,B)$ gravity, models featuring non-linear terms like $f(T,B)=T+\alpha\sqrt{B}$ have been analyzed to understand dark energy and late-time acceleration, with the boundary term $B$ ensuring compatibility with $f(R)$ gravity \cite{Baha23,Apa21}. The $f(T,\tau)$ gravity framework, with $\tau$ denoting the trace of the energy-momentum tensor, has employed forms like $f(T,\tau)=T+\beta\sqrt{\tau}$ to investigate the interplay between matter and geometry and potential violations of energy conditions \cite{LKD24}. In $f(Q,T)$ gravity, which utilizes the symmetric teleparallel formulation, functions like $f(Q,T)=Q+\eta\sqrt{T}$ have been used to investigate the coupling between the non-metricity scalar and matter and its potential role in driving cosmic acceleration \cite{Xu19,Alam2024}. These models inspire the use of square-root and non-linear scalar functions in our $f(T,T_{G})$ model, with the goal of unifying cosmic evolution at both early and late times.
\section{The exponential $Om(z)$ diagnostic and its interpretation in $f(T,T_{G})$ gravity model}\label{sec4}
\hspace{0.5cm} Over the past few years, the $Om(z)$ diagnostic has become a useful cosmological tool for distinguishing among the $\Lambda$CDM model and other theories that involve gravity modifications or evolving dark energy. The $Om(z)$ function differs from conventional methods tied to the equation of state parameter $\omega(z)$, as it provides a geometry-based, model-independent analysis rooted in direct Hubble parameter measurements, bypassing assumptions about dark energy's behavior. To further enhance the diagnostic's capabilities, we introduce a novel parametrization of the $Om(z)$ function, utilizing an exponential redshift dependence:
\begin{equation}\label{21}
Om(z)=\alpha e^{\frac{z}{1+z}}+\beta,
\end{equation}
where $\alpha$ and $\beta$ are free parameters. Unlike polynomial or CPL parametrizations, our form introduces a gradual, saturating evolution with redshift, governed by the $\frac{z}{1+z}$ function, which remains well-behaved at all redshifts from $z=0$ to high redshift $z\rightarrow\infty$. The use of this model is motivated by several properties: $(i)$ The function $\frac{z}{1+z}$ smoothly interpolates between early and late-time epochs as $\frac{z}{1+z}\rightarrow0$ as $z\rightarrow0$ and $\frac{z}{1+z}\rightarrow1$ as $z\rightarrow\infty$. $(ii)$ The exponential enhancement increases the model's sensitivity to small deviations in cosmic expansion while avoiding divergence and $(iii)$ It enables a flexible differentiation between quintessence-like and phantom-like models, characterized by $\omega>-1$ and $\omega<-1$ respectively, without requiring a specific equation of state formulation. 

At low redshift $(z<<1)$, expanding $\frac{z}{1+z}\approx z-z^{2}+...$, we obtain:
\begin{equation}\label{22}
Om(z)\approx \alpha(1+z+...)+\beta,
\end{equation}
The slope at low redshift is predominantly governed by $\alpha$, offering a window into the current-era dynamics of dark energy or modified gravity.

At high redshift $(z>>1)$, $\frac{z}{1+z}\rightarrow1$, the function approaches:
\begin{equation}\label{23}
Om(z)\rightarrow \alpha e+\beta,
\end{equation}
The asymptotic behavior in the matter-dominated era is defined, tracing the model's departure from standard cosmology in the early Universe. A positive $\alpha$ leads to an decreasing $Om(z)$, suggesting quintessence-like dynamics, while a negative $\alpha$ results in a increasing $Om(z)$, consistent with phantom-like behavior; meanwhile, $\alpha=0$, $\beta\approx\Omega_{m0}$ connects the model directly to the present-day matter density.

From the Hubble-parameter-based definition of the $Om(z)$ diagnostic:
\begin{equation}\label{24}
Om(z)=\frac{\bigg(\frac{H(z)}{H_{0}}\bigg)^{2}-1}{(1+z)^{3}-1},
\end{equation}
This can be rearranged to yield an expression for the Hubble parameter:
\begin{equation}\label{25}
H(z)=H_{0}\sqrt{\bigg(\alpha e^{\frac{z}{1+z}}+\beta\bigg)\big[(1+z)^{3}-1\big]+1}.
\end{equation}
This formulation directly captures deviations from $\Lambda$CDM in the cosmic expansion rate, making it readily comparable to observational data from BAO, cosmic chronometers and SNe Ia (Pantheon+).

To better understand the evolution of cosmic expansion and its departure from the standard model, we derive the time derivatives of the Hubble parameter. These comprise the first derivative $\dot{H}$, the second derivative $\ddot{H}$ and the third derivative $\dddot{H}$ which are crucial for calculating dynamical quantities like the deceleration parameter $q(z)$, energy density $\rho(z)$, pressure $p(z)$ and statefinder parameters $\{r,s\}$. Applying the chain rule, we can compute the cosmic time derivatives of $H(z)$ based on its redshift dependence:
\begin{equation}\label{26}
\dot{H}=-\frac{H_{0}^{2}}{2}\bigg[3\bigg(\alpha e^{\frac{z}{1+z}}+\beta\bigg)(1+z)^{3}+\frac{(1+z)^{3}-1}{(1+z)}\alpha e^{\frac{z}{1+z}}\bigg],
\end{equation}
\begin{eqnarray}\label{27}
\ddot{H}&=&\frac{H_{0}^{3}}{2}\sqrt{\bigg(\alpha e^{\frac{z}{1+z}}+\beta\bigg)\big[(1+z)^{3}-1\big]+1}\times\bigg[3\alpha(1+z)^{2}e^{\frac{z}{1+z}}+9(1+z)^{3}\bigg(\alpha e^{\frac{z}{1+z}}+\beta\bigg)\\\nonumber
&&+\alpha e^{\frac{z}{1+z}}\bigg\{\frac{1+2z^{4}+9z^{3}+14z^{2}+9z}{(1+z)^{2}}\bigg\}\bigg],
\end{eqnarray}
\begin{eqnarray}\label{28}
\dddot{H}&=&-\frac{H_{0}^{4}}{2}\;\bigg\{\bigg(\alpha e^{\frac{z}{1+z}}+\beta\bigg)3(1+z)^{3}+\frac{(1+z)^{3}-1}{(1+z)}\alpha e^{\frac{z}{1+z}}\bigg\}\times\bigg[3\alpha(1+z)^{2}e^{\frac{z}{1+z}}+9(1+z)^{3}\\\nonumber
&&\bigg(\alpha e^{\frac{z}{1+z}}+\beta\bigg)+\alpha e^{\frac{z}{1+z}}\bigg\{\frac{1+2z^{4}+9z^{3}+14z^{2}+9z}{(1+z)^{2}}\bigg\}\bigg]-\frac{H_{0}^{4}}{2}\bigg[\bigg(\alpha e^{\frac{z}{1+z}}+\beta\bigg)\\\nonumber
&&\big[(1+z)^{3}-1\big]+1\bigg]\times\bigg\{15\alpha(1+z)^{2}e^{\frac{z}{1+z}}+3\alpha(1+z)e^{\frac{z}{1+z}}+27(1+z)^{3}\bigg(\alpha e^{\frac{z}{1+z}}+\beta\bigg)\\\nonumber
&&+\alpha e^{\frac{z}{1+z}}\bigg(\frac{1+2z^{4}+9z^{3}+14z^{2}+9z}{(1+z)^{2}}\bigg)+\alpha e^{\frac{z}{1+z}}\bigg(\frac{8z^{3}+27z^{2}+28z+9}{(1+z)}\\\nonumber
&&-\frac{2(1+2z^{4}+9z^{3}+14z^{2}+9z)}{(1+z)^{2}}\bigg)\bigg\}.
\end{eqnarray}
\section{Parameter estimation using observational data}\label{sec5}
\hspace{0.5cm} In this section, we constrain the parameters $H_{0}$, $\alpha$ and $\beta$ of our cosmological model by employing a range of observational datasets. The analysis is based on a maximum likelihood approach, where the total likelihood is defined as $\mathcal{L}\propto e^{-\frac{\chi^{2}}{2}}$ and the total chi-square is constructed as the sum of contributions from various independent sources:
\begin{equation}\label{29}
\chi^{2}_{total}=\chi^{2}_{CC}+\chi^{2}_{BAO}+\chi^{2}_{SNe Ia},
\end{equation}
To determine the optimal estimates for $H_{0}$, $\alpha$ and $\beta$, we employ a Markov Chain Monte Carlo (MCMC) method. This approach systematically samples the parameter space to identify values that minimize the total chi-square statistic. Parameter estimation is conducted by constraining $H_{0}$, $\alpha$ and $\beta$ within the ranges $[60,100]$, $[-1,1]$ and $[0,1]$, respectively.
\subsection{Cosmic Chronometers (CC)}\label{sec5.1}
\hspace{0.5cm} The CC technique offers a model-independent way to estimate the Hubble parameter $H(z)$ by utilizing the age differences of massive, passively evolving galaxies. This method relies on measuring their relative ages to reconstruct the expansion rate of the Universe. By looking at galaxy pairs with small redshift differences, one can measure the differential age $\frac{dz}{dt}$ and directly link it to the Hubble parameter as: $H(z)=-\frac{1}{1+z}\frac{dz}{dt}$ \cite{Moresco16,AS25}. We employ a dataset of $31$ CC points, covering redshifts from $0.07$ to $2.41$, based on the compilation by \cite{NM23}. We compute the chi-square statistic to quantify differences between modeled and observed Hubble parameters:
\begin{equation}\label{30}
\chi^{2}_{CC}=\sum_{j=1}^{31}\frac{\big[H(z_{j},H_{0},\alpha,\beta)-H_{obs}(z_{j})\big]^{2}}{\sigma^{2}(z_{j})},
\end{equation}
In this context, $H_{0},\alpha,\beta$ represents the model-predicted Hubble parameter, whereas $H_{obs}(z_{j})$ and its uncertainty $\sigma^{2}(z_{j})$ refers to the empirical data points and their respective uncertainties at the redshift $z_{j}$.
\subsubsection{BAO Measurements}\label{sec5.2}
\hspace{0.5cm} The BAO signal acts as a fixed reference scale that astronomers use to track the Universe's expansion history. These oscillations are a result of sound waves in the early photon-baryon plasma, leaving a signature on the large-scale structure of galaxy distributions. The latest BAO measurements from galaxies, quasars, and the Lyman-$\alpha$ forest, reported by the DESI collaboration, are employed in our study \cite{DESI24,AG25,AGA25}. The following quantities are central to our analysis:
\begin{equation}\label{31}
\text{Transverse comoving distance ratio:}\hspace{0.4cm} \frac{d_{M}(z)}{r_{d}}=\frac{D_{L}(z)}{r_{d}(1+z)},
\end{equation}
\begin{equation}\label{32}
\text{Hubble distance ratio:}\hspace{0.4cm} \frac{d_{H}(z)}{r_{d}}=\frac{c}{r_{d}H(z)}, 
\end{equation}
\begin{equation}\label{33}
\text{Volume-averaged distance ratio:}\hspace{0.4cm} \frac{d_{V}(z)}{r_{d}}=\frac{\big[zd_{M}^{2}(z)d_{H}(z)\big]^{\frac{1}{3}}}{r_{d}},
\end{equation}
The drag epoch sound horizon $r_{d}$ is calculated according to the procedure detailed in \cite{DJ05}. The BAO chi-square is calculated as:
\begin{equation}\label{34}
\chi^{2}_{BAO}=\sum_{j=1}^{N} \bigg[\frac{Y_{j}^{th}-Y_{j}^{obs}}{\sigma_{Y_{i}}}\bigg]^{2},
\end{equation}
where $Y_{j}^{th}$ are the theoretical BAO observables from our model and $Y_{j}^{obs}$ are the corresponding measurements.
\subsubsection{Pantheon+ Supernovae data}\label{sec5.3}
\hspace{0.5cm} Type Ia supernovae are valuable cosmological probes due to their consistent maximum luminosity, allowing for precise distance calculations. We base our analysis on the Pantheon+ compilation, which contains $1701$ light curves from $1550$ SNe Ia, with redshifts extending between $z = 0.001$ to $z = 2.26$ \cite{Kowa08}-\cite{Scolnic18}. The theoretical distance modulus is given by:
\begin{equation}\label{35}
\mu_{th}(z; \Theta) = 5 \log_{10} \left[ d_L(z; \Theta) \right] + 25,
\end{equation}
where the luminosity distance $d_{L}(z)$ is computed as:
\begin{equation}\label{36}
d_L(z; \Theta) = c (1 + z) \int_0^z \frac{dz'}{H(z'; H_0, \alpha, \beta)}.
\end{equation}
The chi-square for the Pantheon+ dataset is:
\begin{equation}\label{37}
\chi^2_{SNe Ia} = \sum_{j=1}^{1701} \left( \frac{\mu_{obs}(z_j) - \mu_{th}(z_j; \Theta)}{\sigma_{\mu}(z_j)} \right)^2,
\end{equation}
where $\mu_{obs}$ and $\sigma_{\mu}(z_j)$ are the observed distance modulus and its uncertainty, respectively.
\subsection{Joint Constraints and Confidence Intervals}\label{sec5.4}
\begin{figure}[hbt!]
  \centering
  \includegraphics[scale=0.33]{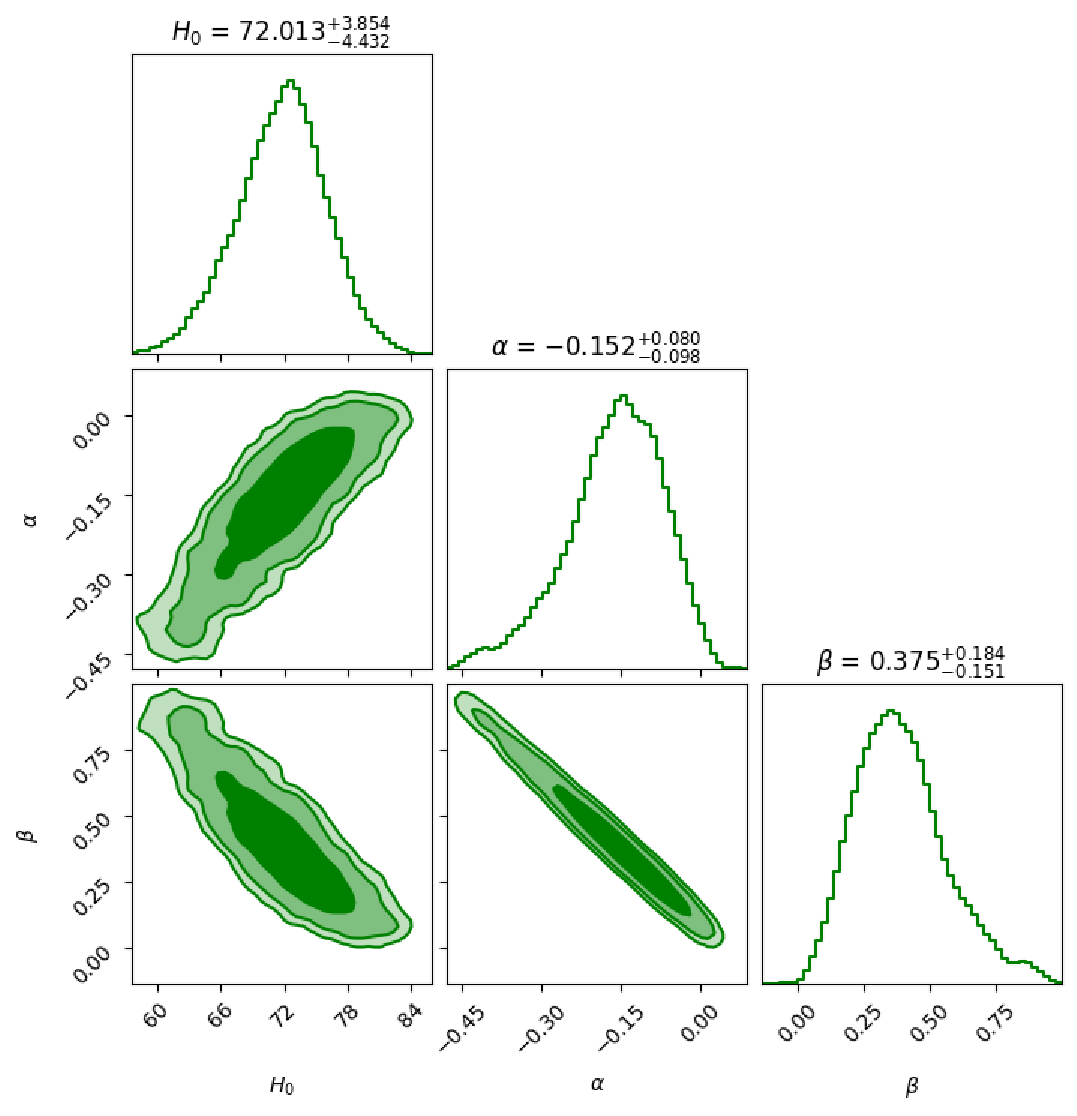}~~~
  \includegraphics[scale=0.33]{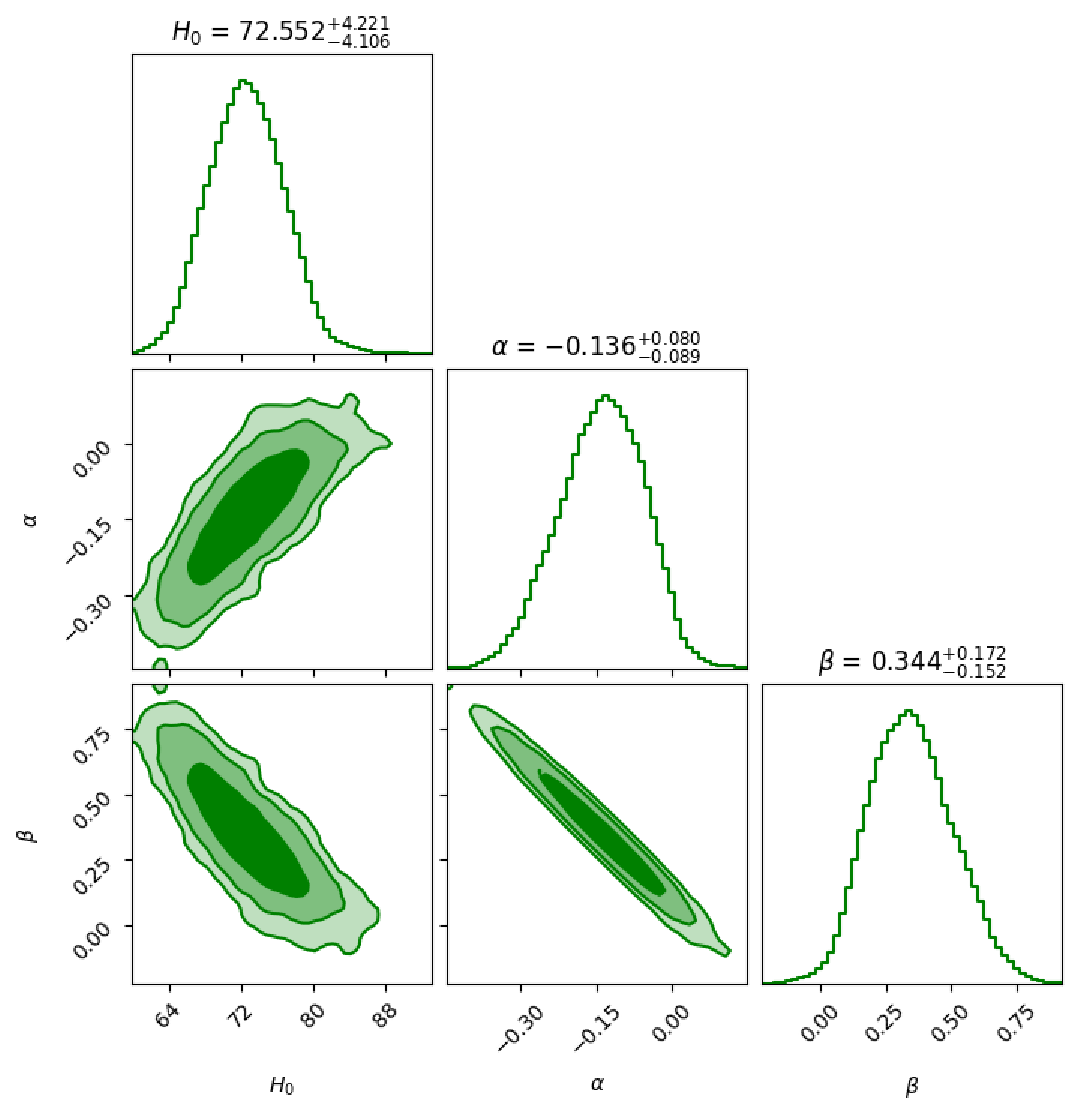}\\
  \includegraphics[scale=0.33]{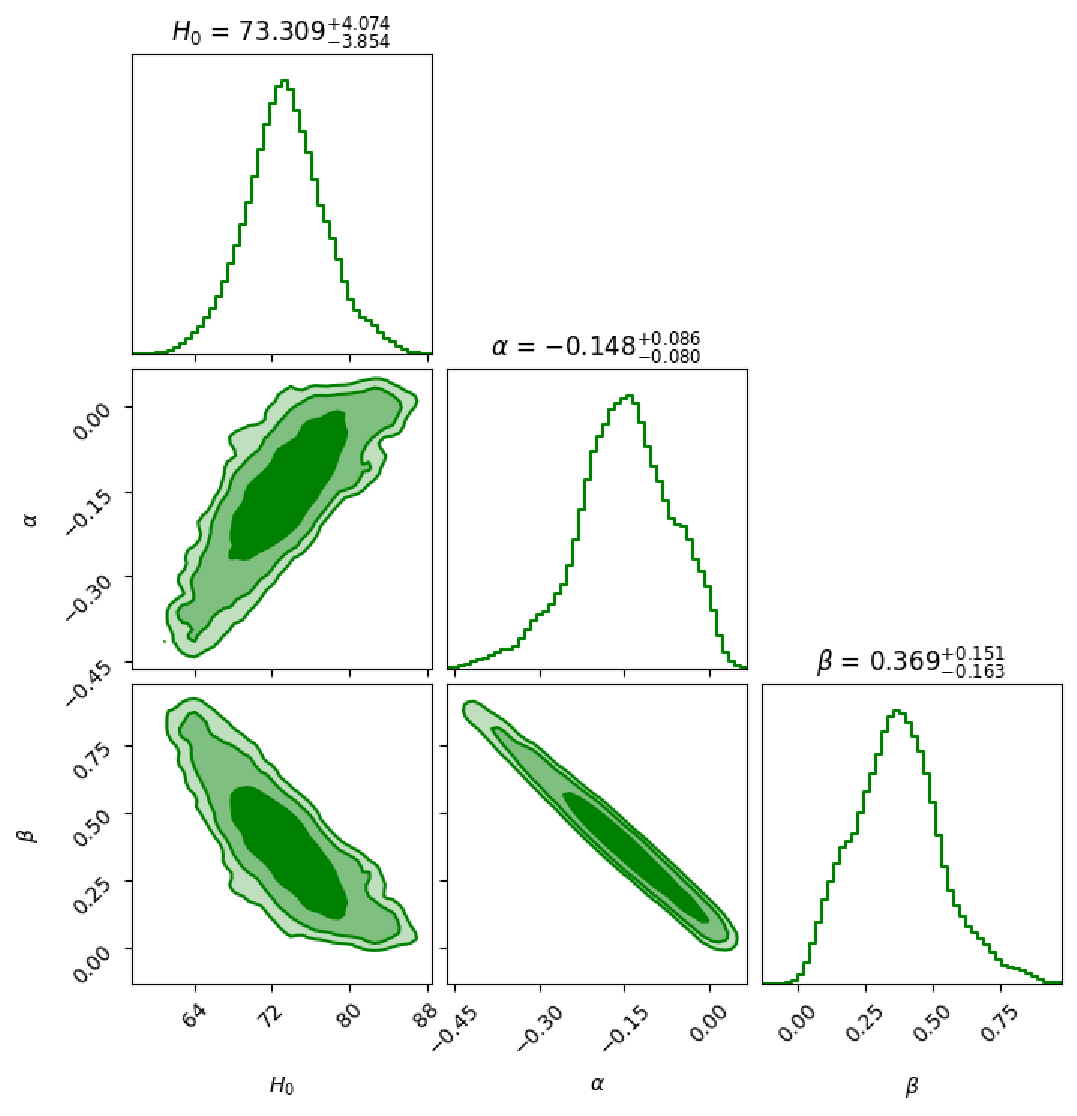}
  \caption{Joint confidence regions for $(H_{0}, \alpha, \beta)$ parameters from various dataset combinations. The shaded regions denote the $1\sigma$ $(68.27\%)$, $2\sigma$ $(95.45\%)$ and $3\sigma$ $(99.73\%)$ confidence levels.}\label{fig:f1}
\end{figure}

\begin{figure}[hbt!]
  \centering
  \includegraphics[scale=0.3]{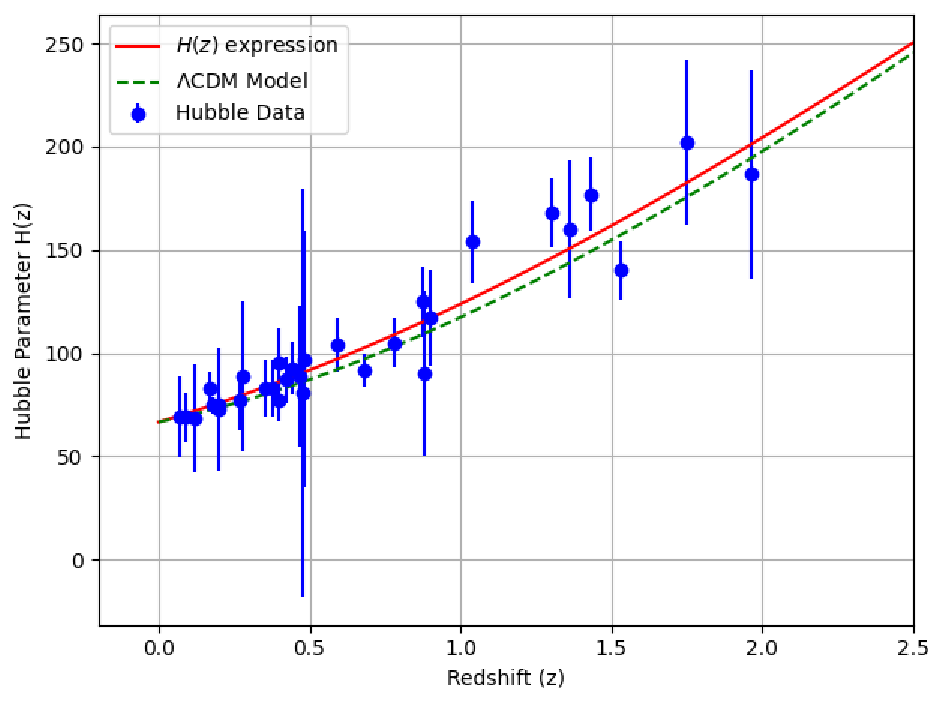}~
  \includegraphics[scale=0.33]{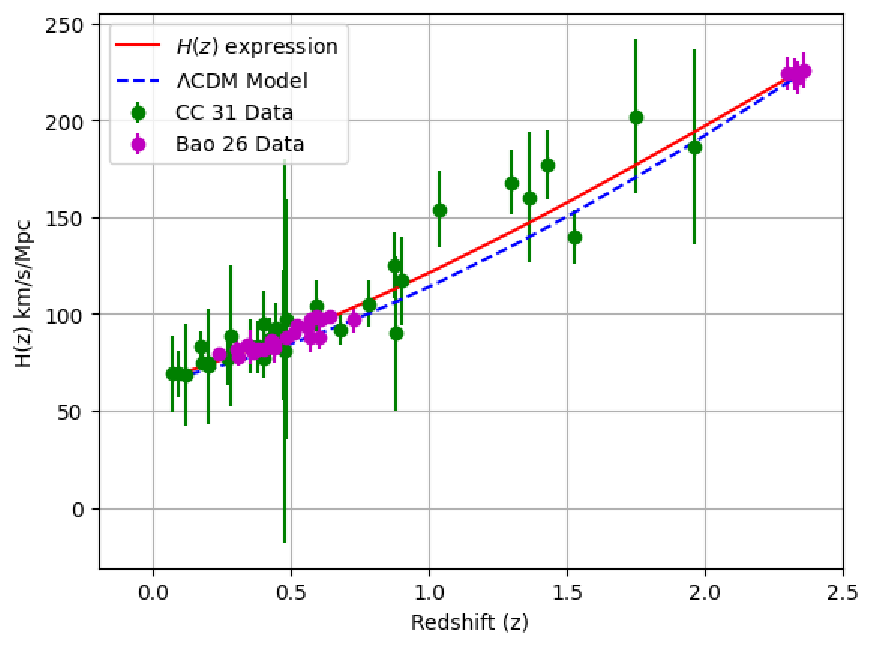}~~
  \includegraphics[scale=0.28]{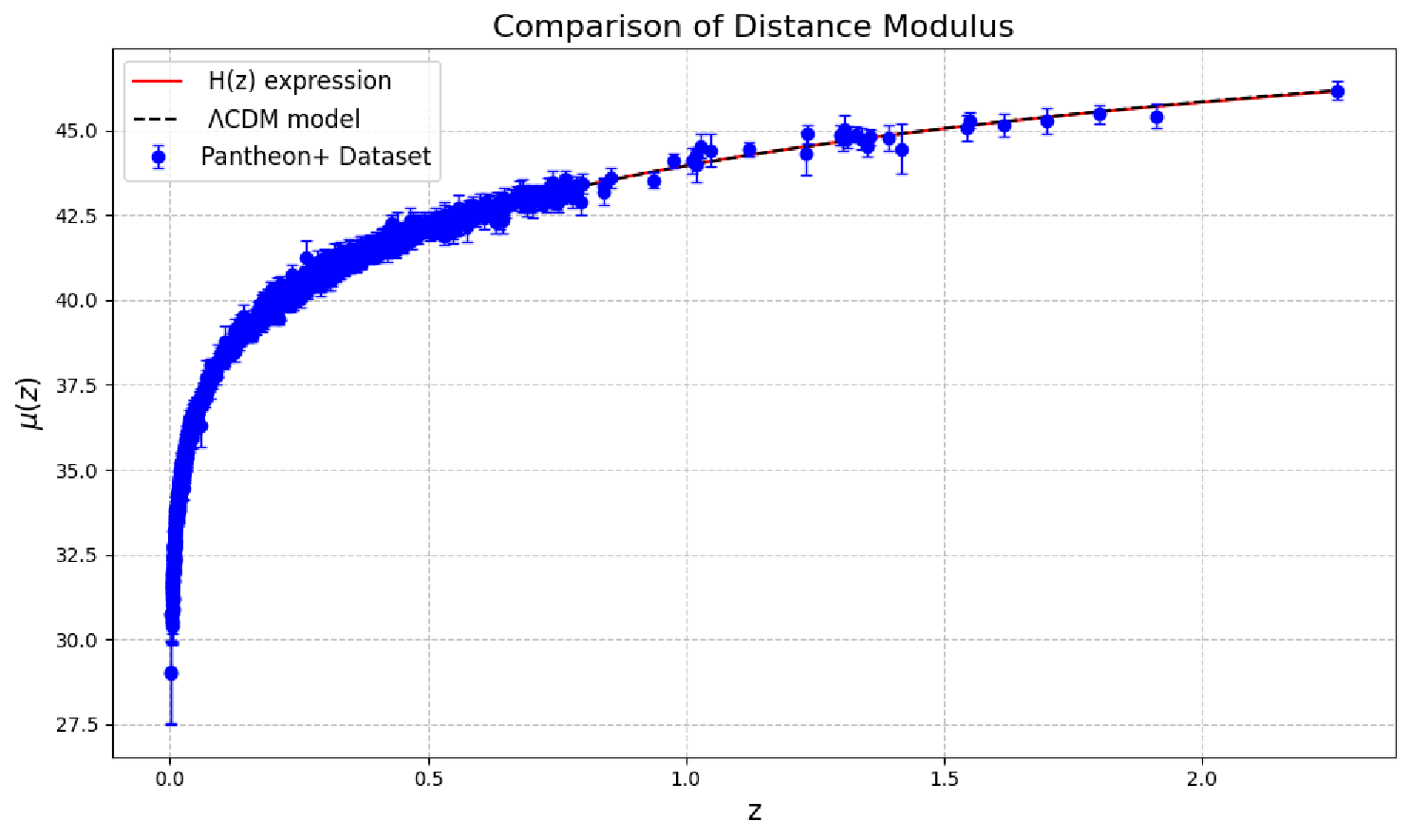}
  \caption{Error bar analysis depicting variations in parameter estimates from dataset combinations.}\label{fig:f2}
\end{figure}

\hspace{0.5cm} The full statistical model combines constraints from CC, BAO and SNe Ia data into a single chi-square function:
\begin{equation}\label{38}
\chi^{2}_{total}=\chi^{2}_{CC}+\chi^{2}_{BAO}+\chi^{2}_{SNe Ia},
\end{equation}
Using this cumulative analysis, we generate confidence level contours for parameter pairs such as $(H_{0},\alpha)$, $(\alpha,\beta)$ and $(H_{0},\beta)$ demonstrating how multi-probe observational data narrows down the viable parameter space. Figures \ref{fig:f1} and \ref{fig:f2} illustrate these constraints and the evolution of observational uncertainty across datasets.

Our analysis constrained our model parameters $H_{0}$, $\alpha$ and $\beta$ with three dataset combinations: CC alone, CC+BAO and CC+BAO+Pantheon+, with key findings summarized as follows:\\
$\bullet$ CC: $H_{0}=72.013^{+3.854}_{-4.432}$ km/s/Mpc, $\alpha=-0.152^{+0.080}_{-0.098}$ and $\beta=0.375^{+0.184}_{-0.151}$, \\
$\bullet$ CC+BAO: $H_{0}=72.552^{+4.221}_{-4.106}$ km/s/Mpc, $\alpha=-0.136^{+0.080}_{-0.089}$ and $\beta=0.344^{+0.172}_{-0.152}$, \\
$\bullet$ CC+BAO+Pantheon+: $H_{0}=73.309^{+4.074}_{-3.854}$ km/s/Mpc, $\alpha=-0.148^{+0.086}_{-0.080}$ and $\beta=0.369^{+0.151}_{-0.163}$.\\
Our Hubble constant estimates from all three cases align with the local measurement of $H_{0}=73.04\pm 1.04$ km/s/Mpc from \cite{Adam21}, indicating our model may help resolve the discrepancy between early and late-Universe Hubble measurements. The parameters $\alpha$ and $\beta$ govern the deviation of the model from the standard $\Lambda$CDM. A negative value of $\alpha$ indicates a dynamical evolution in the $Om(z)$ function, pointing toward a deviation from a constant dark energy behavior. The positive value of $\beta$ controls the curvature or redshift-dependence of $Om(z)$, with increasing $\beta$ values suggesting a more pronounced deviation from $\Lambda$CDM at higher redshifts. Given the negative $\alpha$ and positive $\beta$, the behavior of our $Om(z)$ model indicates: $(i)$ A slowly increasing function with redshift, i.e., $Om(z)$ increases with $z$, which suggests a phantom-like behavior in the dark energy sector $(\omega<-1)$. $(ii)$ This implies that the dark energy density might grow over time, potentially leading to a scenario with future singularities or rapid acceleration, depending on how the parameters evolve further. These results indicate that our proposed exponential form of the $Om(z)$ model remains fully consistent with all current observational constraints from CC, BAO and Pantheon+ data, thereby providing a viable and observationally supported framework for describing the late-time cosmic expansion.
\section{Behavior of cosmological parameters}\label{sec6}
\subsection{Deceleration parameter}\label{sec6.1}
\hspace{0.5cm} To further investigate the implications of our observational constraints, we now analyze the behavior of key cosmological parameters derived from the model. The deceleration parameter $q(z)$ is our starting point, providing a characterization of cosmic acceleration and insight into the Universe's transition from a decelerated to an accelerated expansion era. The deceleration parameter is derived for our model as:
\begin{equation}\label{39}
q=-1-\frac{\dot{H}}{H^{2}}=-1+\frac{3\bigg(\alpha e^{\frac{z}{1+z}}+\beta\bigg)(1+z)^{3}+\frac{(1+z)^{3}-1}{(1+z)}\alpha e^{\frac{z}{1+z}}}{\bigg(\alpha e^{\frac{z}{1+z}}+\beta\bigg)\big[(1+z)^{3}-1\big]+1}.
\end{equation}
The sign of $q(z)$ determines the Universe's expansion state: negative for acceleration and positive for deceleration. This parameter is especially useful for identifying the redshift $z_{t}$ at which the universe transitioned from deceleration to acceleration. By substituting the best-fit values of $\alpha$ and $\beta$ from our observational analysis into the reconstructed Hubble parameter $H(z)$, we examine the evolution of $q(z)$ and assess whether it aligns with the expected cosmic history suggested by current data.
\begin{figure}[hbt!]
  \centering
  \includegraphics[scale=0.4]{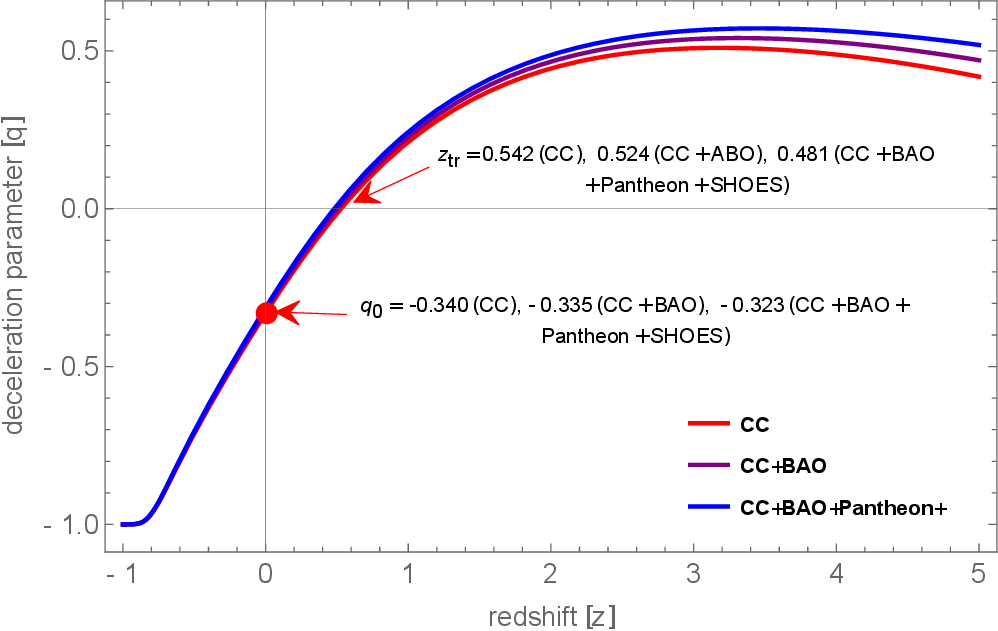}
  \caption{$q$ vs. $z$ plot}\label{fig:f3}
\end{figure}

Figure \ref{fig:f3} shows how the deceleration parameter $q(z)$ evolves according to our model, using parameters obtained from different dataset combinations. The plot illustrates a gradual transition from an early universe with $q>0$ (deceleration) to a current state with $q<0$ (acceleration), in line with cosmological predictions. The redshift at which this transition occurs, denoted as $z_{tr}$, shows a systematic shift as more datasets are included: we find $z_{tr}=0.542$ for CC data, $z_{tr}=0.524$ for CC+BAO and $z_{tr}=0.481$ for CC+BAO+Pantheon+ dataset. This decreasing trend in $z_{tr}$ indicates that the acceleration phase is realized earlier in cosmic history when more comprehensive data are considered. The current deceleration parameter, $q_{0}$, becomes increasingly negative across different datasets: CC: $q_{0}=-0.340$, CC+BAO: $q_{0}=-0.335$ and CC+BAO+Pantheon+: $q_{0}=-0.323$. Our results align with current observational constraints, demonstrating that our model accurately captures the Universe's expansion: an early deceleration phase followed by a late-time acceleration, consistent with established cosmological behavior.
\subsection{Energy density and pressure}\label{sec6.2}
\hspace{0.5cm} In the study of cosmology, energy density $\rho$ and pressure $p$ play a central role in shaping the Universe's expansion dynamics. Their evolution determines whether the Universe is accelerating or decelerating. For our specific $F(T,T_{G})$ model, the effective energy density and pressure are derived using the modified Friedmann equations. These are computed through the Hubble parameter $H(z)$ and its derivatives, as given in equations (\ref{25})–(\ref{28}). The explicit forms reflect how the modified gravity terms contribute to the overall energy-momentum budget.
\begin{equation}\label{40}
\rho=-3H^{2}+\frac{\gamma\sqrt{6}H}{2}\sqrt{\dot{H}+H^{2}}-2\sqrt{\frac{3}{2}}\gamma H\bigg[\frac{2\dot{H}^{2}+H\ddot{H}+4\dot{H}H^{2}}{(\dot{H}+H^{2})^{\frac{3}{2}}}\bigg],
\end{equation}
\begin{eqnarray}\label{41}
p&=&3H^{2}-\frac{\gamma\sqrt{6}H}{2}\sqrt{\dot{H}+H^{2}}+2\dot{H}+4\gamma\sqrt{6}H\sqrt{\dot{H}+H^{2}}\;(2\dot{H}^{2}+H\ddot{H}+4\dot{H}H^{2})-\frac{\gamma}{4}\sqrt{\frac{3}{2}}\\\nonumber
&&\frac{1}{(\dot{H}+H^{2})^{\frac{5}{2}}}(2\dot{H}^{2}+H\ddot{H}+4\dot{H}H^{2})^{2}+\frac{\gamma}{\sqrt{24}H^{3}(\dot{H}+H^{2})^{\frac{3}{2}}}(2\dot{H}^{3}+6H\dot{H}\ddot{H}+\dddot{H}H^{2}\\\nonumber
&&+8H^{2}\dot{H}^{2}+4H^{3}\ddot{H}).
\end{eqnarray}
\begin{figure}[hbt!]
  \centering
  \includegraphics[scale=0.4]{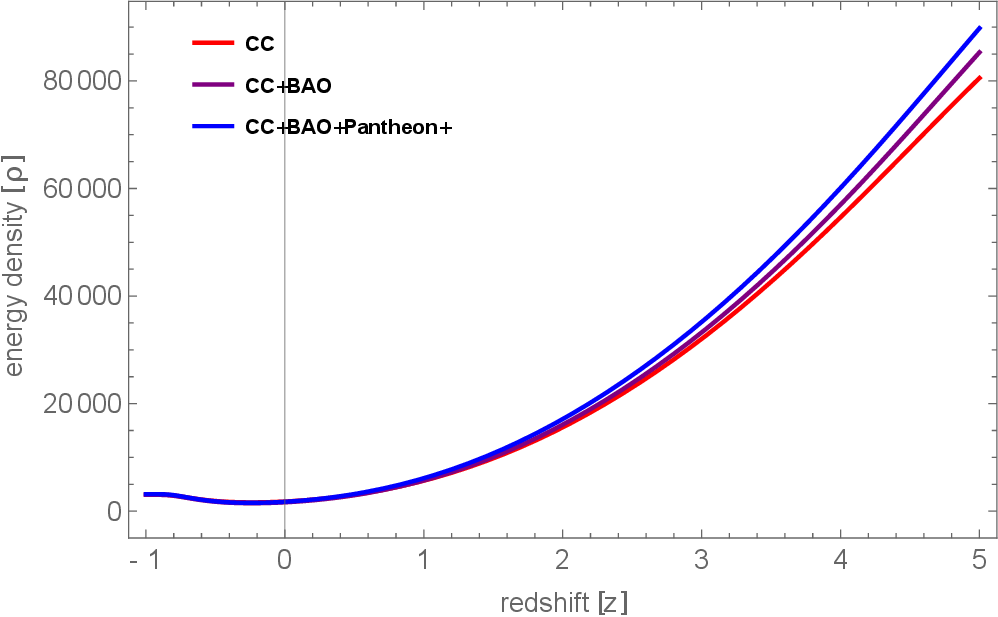}~~~
  \includegraphics[scale=0.4]{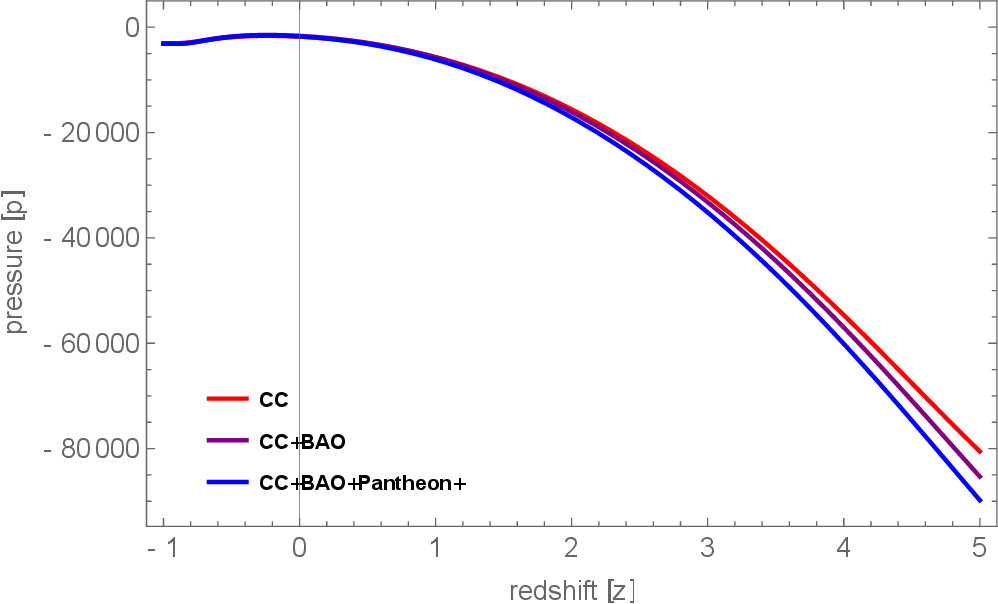}
  \caption{Energy density and pressure vs. redshift for the parameter $\gamma=0.05$.}\label{fig:f4}
\end{figure}

Figure \ref{fig:f4} illustrates the behavior of energy density $\rho$ and pressure $p$ as a function of redshift, as predicted by our model. From the plots, it is evident that the energy density remains positive throughout cosmic history, ensuring physical viability and compatibility with observations of matter and dark energy contributions. The pressure transitions to and remains negative at late times, which is essential for driving the accelerated expansion of the Universe. These features are fully consistent with current observational data, such as Type Ia supernovae, CMB and large-scale structure measurements and align with the standard cosmological requirement for dark energy-like behavior. The results support the reliability of the $f(T,T_{G})$ model in explaining late-time cosmic acceleration while preserving theoretical consistency with established cosmological principles.
\subsection{EoS parameter}\label{sec6.3}
\hspace{0.5cm} The EoS parameter $\omega(z)=\frac{p(z)}{\rho(z)}$ plays a crucial role in characterizing the nature of cosmic fluids driving the expansion of the Universe. In standard cosmology: $\omega=0$ corresponds to pressureless matter (dust), $\omega=0.33$ represents radiation, $\omega=-1$ mimics a cosmological constant (dark energy), $\omega<-1$ indicates a phantom-like behavior and $-1<\omega<-0.33$ corresponds to quintessence-like dark energy. The EoS parameter in our $F(T,T_{G})$ gravity model is determined from the energy density and pressure relations specified in equations (\ref{40}) and (\ref{41}). These expressions incorporate the Hubble parameter and its derivatives, thus encapsulating the impact of modified gravity on cosmic dynamics.
\begin{equation}\label{42}
\omega=-1-\frac{-2\dot{H}f_{T}-2H\dot{f}_{T}+\frac{T_{G}\dot{f}_{T_{G}}}{3H}+4H^{2}\ddot{f}_{T_{G}}-12H^{3}\dot{f}_{T_{G}}}
{\frac{f}{2}-6H^{2}f_{T}-\frac{T_{G}f_{T_{G}}}{2}+12H^{3}\dot{f}_{T_{G}}}.
\end{equation}
\begin{figure}[hbt!]
  \centering
  \includegraphics[scale=0.4]{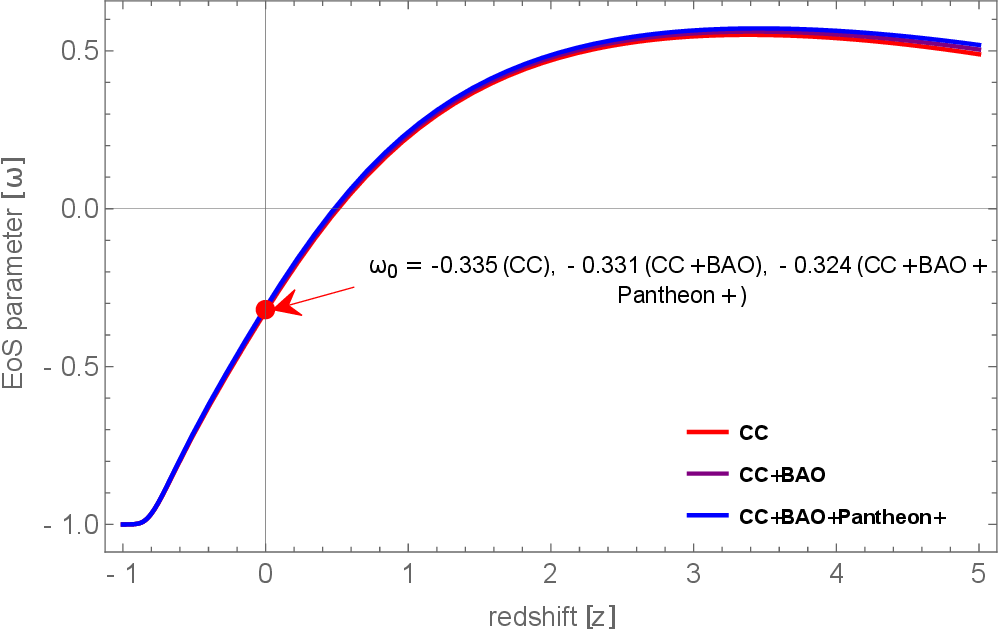}
  \caption{EoS parameter vs. redshift for the parameter $\gamma=0.05$.}\label{fig:f5}
\end{figure}

Figure \ref{fig:f5} illustrates how the equation of state parameter $\omega(z)$ evolves with redshift across different datasets. From the plot, we observe that: at higher redshift $(z>1)$, $\omega(z)\approx0.5$ resembling a stiff matter or radiation-like behavior. As redshift decreases, $\omega(z)$ steadily declines, crossing into the dark energy regime. At present time $(z=0)$, the EoS values are: $\omega_{0}=-0.335$ (CC), $\omega_{0}=-0.331$ (CC+BAO) and $\omega_{0}=-0.324$ (CC+BAO+Pantheon+). These values lie well within the quintessence regime and closely approach $\omega=-1$, ndicating a late-time accelerated expansion without invoking phantom energy. The smooth transition from positive to negative values of $\omega$ reflects a dynamically evolving dark energy component. Such behavior is compatible with current observational constraints from Planck, SNe Ia, and BAO data, which generally favor a time-evolving EoS slightly above $\omega=-1$. Thus, our model not only explains the late-time acceleration but does so in a manner consistent with observational expectations \cite{Amit24,CA23}.
\subsection{Analysis of the $Om(z)$ function}\label{sec6.4}
\begin{figure}[hbt!]
  \centering
  \includegraphics[scale=0.4]{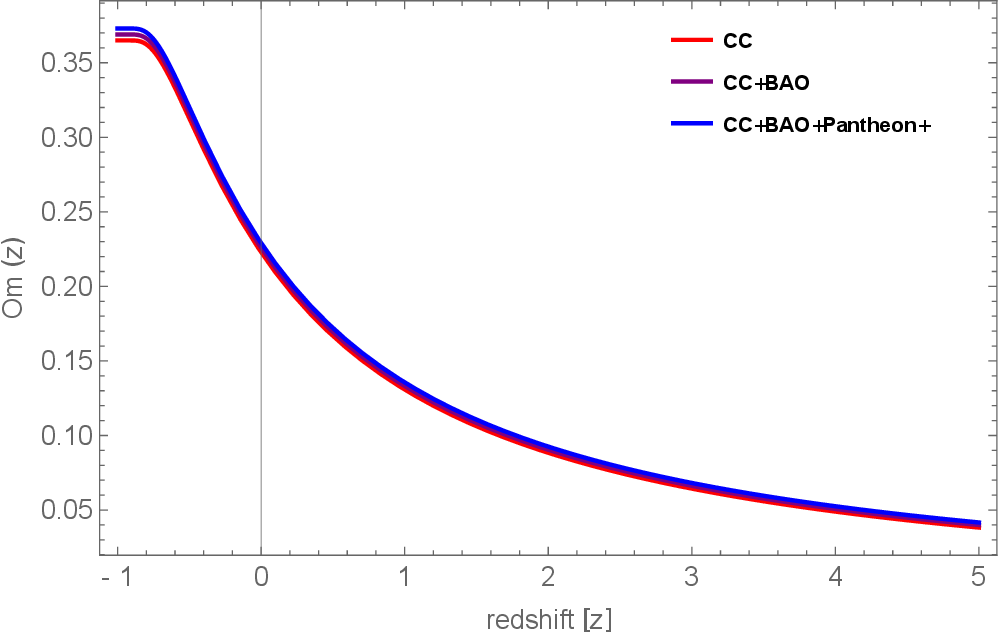}
  \caption{$Om(z)$ vs. redshift.}\label{fig:f6}
\end{figure} 

The plot of $Om(z)$ versus redshift shows a clear increasing trend, starting near $0.05$ at higher redshifts and approaching approximately $0.36$ as $z\rightarrow-1$. This behavior is consistent with a dynamical dark energy scenario and reflects a deviation from the constant $Om(z)$ expected in $\Lambda$CDM. The negative values of $\alpha$ and positive $\beta$ drive this evolution, supporting a phantom-like EoS $(\omega<-1)$. This indicates that the dark energy density increases with time, potentially pointing toward a future-accelerated expansion. The behavior of the curve, along with the observationally constrained parameter ranges, confirms that our exponential $Om(z)$ model is compatible with current data and successfully describes the late-time acceleration phase of the Universe.
\section{Energy conditions in $f(T,T_{G})$ gravity}\label{sec7}
\hspace{0.5cm} Energy conditions serve as essential tools in gravitational theories to assess the physical plausibility of solutions, particularly in relation to geodesic focusing, causal structure and the viability of matter content. These conditions originate from general relativity but are frequently analyzed in extended theories such as $f(T,T_{G})$ gravity to understand how modified geometrical terms influence cosmic evolution. The standard energy conditions are defined as follows: $(i)$ Null Energy Condition (NEC): $\rho+p\geq0$, $(ii)$ Dominant Energy Condition (DEC): $\rho-p\geq0$ and $(iii)$ Strong Energy Condition (SEC): $\rho+3p\geq0$. These conditions are evaluated using the energy density $\rho$ and pressure $p$, derived from the modified field equations—in our case, from equations (\ref{40}) and (\ref{41}). 
\begin{equation}\label{43}
\rho+p=2\dot{H}f_{T}+2H\dot{f}_{T}-\frac{T_{G}\dot{f}_{T_{G}}}{3H}-4H^{2}\ddot{f}_{T_{G}}+12H^{3}\dot{f}_{T_{G}}\geq0,
\end{equation}
\begin{equation}\label{44}
\rho-p=f-12H^{2}f_{T}-T_{G}F_{T_{G}}+12H^{3}\dot{f}_{T_{G}}-2\dot{H}f_{T}-2H\dot{f}_{T}+\frac{T_{G}\dot{f}_{T_{G}}}{3H}+4H^{2}\ddot{f}_{T_{G}}\geq0,
\end{equation}
\begin{equation}\label{45}
\rho+3p=-f+12H^{2}f_{T}+T_{G}f_{T_{G}}+12H^{3}\dot{f}_{T_{G}}+6\dot{H}f_{T}+6H\dot{f}_{T}-\frac{T_{G}\dot{f}_{T_{G}}}{H}-12H^{2}\ddot{f}_{T_{G}}\geq0.
\end{equation}
\begin{figure}[hbt!]
  \centering
  \includegraphics[scale=0.4]{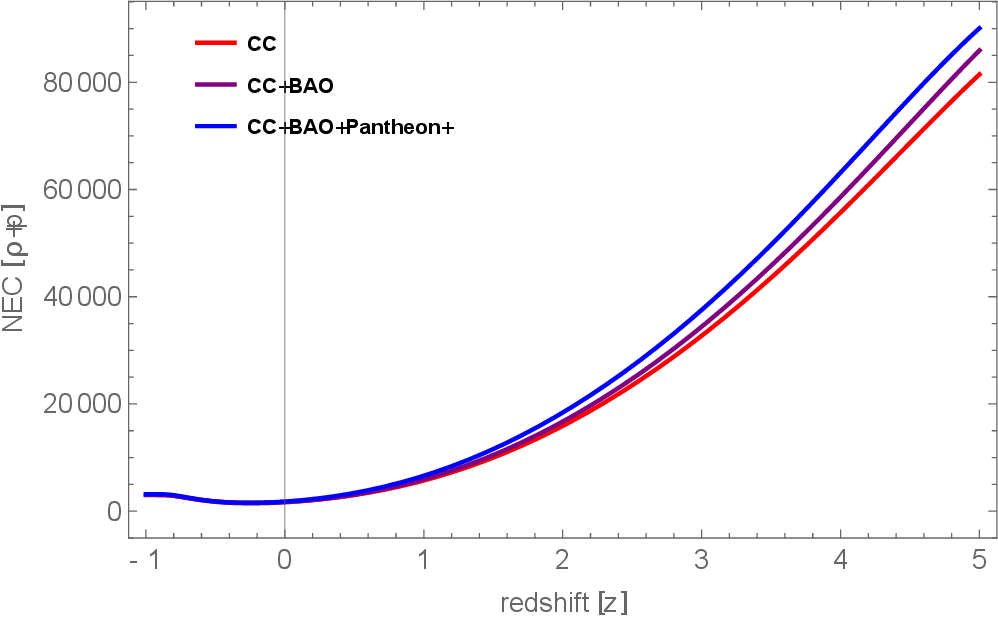}~
  \includegraphics[scale=0.4]{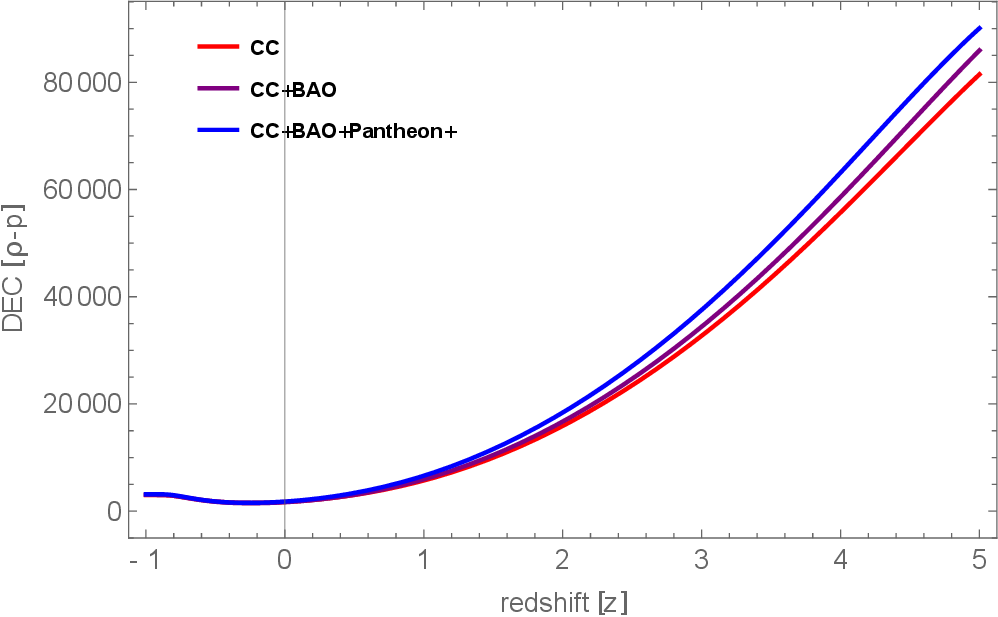}\\
  \includegraphics[scale=0.4]{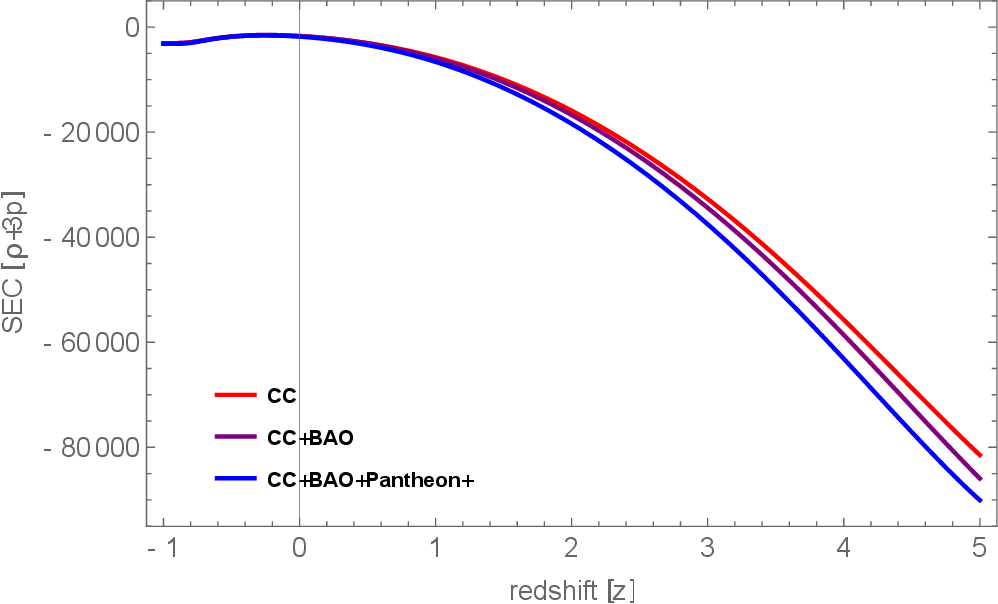}
  \caption{Redshift evolution of energy conditions for the parameter $\gamma=0.05$.}\label{fig:f7}
\end{figure}

In Figure \ref{fig:f7}, we depict the energy conditions' evolution, calculated using the expressions from our model. From the analysis of the figure, we find the following: the NEC and DEC are satisfied throughout the cosmic evolution and the SEC is violated at late times. The violation of the strong energy condition is not a flaw but rather a crucial feature of models that accommodate the current accelerated phase of the Universe. In GR and modified gravity, a violation of SEC is often associated with a repulsive gravitational effect that is exactly what is required to explain the observed cosmic acceleration without invoking exotic matter. Hence, the behavior of the energy conditions in our $f(T,T_{G})$ model aligns well with modern cosmological observations and the theoretical expectations from both GR and extended gravity frameworks. This further validates the model’s capability to describe the Universe’s expansion history in a physically consistent manner \cite{Sam24}.
\section{Statefinder diagnostic and its cosmological implications}\label{sec8}
\hspace{0.5cm} The $\{r, s\}$ statefinder diagnostic serves as a higher-order geometrical probe to differentiate between various dark energy and modified gravity scenarios. In contrast to the Hubble and deceleration parameters, which depend on lower-order derivatives, the statefinder parameters utilize higher-order derivatives to provide a more detailed classification of expansion dynamics. The statefinder pair is defined as \cite{Sahni03}:
\begin{equation}\label{46}
r=\frac{\dddot a}{aH^{3}}=2q^{2}+q+(1+z)\frac{dq}{dz},
\end{equation}
\begin{equation}\label{47}
s=\frac{(r-1)}{3(q-\frac{1}{2})}. \bigg(q\neq\frac{1}{2}\bigg).
\end{equation}
In the standard $\Lambda$CDM model, the statefinder values are fixed at: $\{r, s\}=(1,0)$, which serves as a reference point in the $\{r, s\}$ plane. Deviations from this point help to characterize alternative models like quintessence, phantom, or Chaplygin gas models.
\begin{figure}[hbt!]
  \centering
  \includegraphics[scale=0.4]{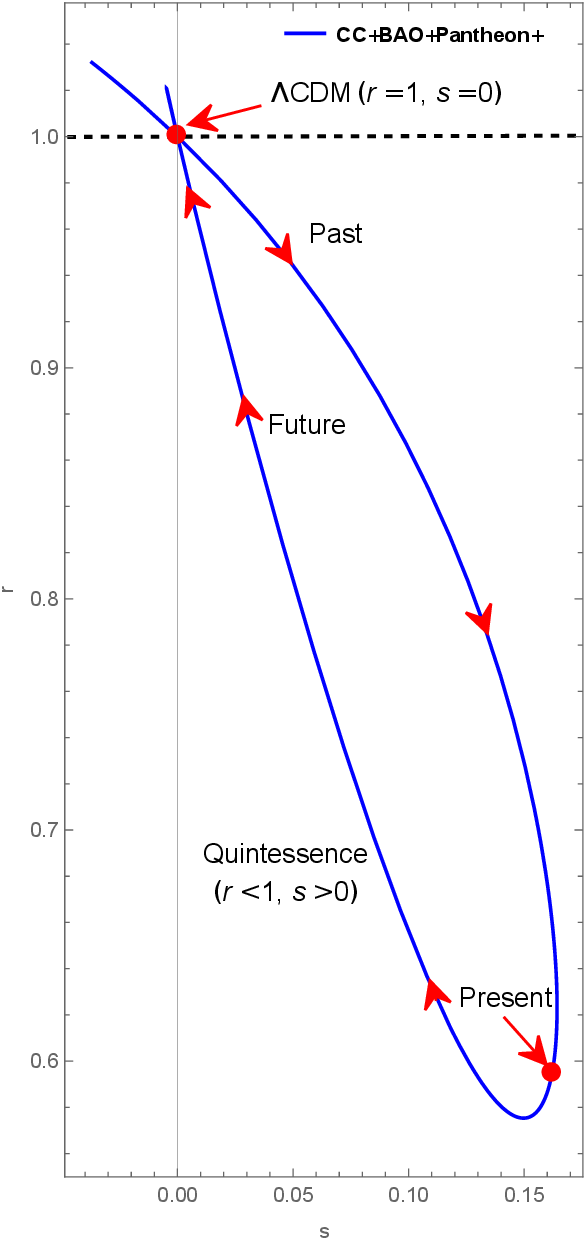}
  \caption{Statefinder trajectory in the $\{r, s\}$ plane for the CC+BAO+Pantheon+ dataset.}\label{fig:f8}
\end{figure}

Figure \ref{fig:f8} shows the trajectory of our model in the $\{r, s\}$ plane for the combined CC+BAO+Pantheon+ dataset. The trajectory clearly passes through the $\Lambda$CDM point $(r=1,s=0)$, indicating that the model is consistent with the standard cosmological model at the present epoch. In the past, the curve lies in the quintessence region, where $r<1$ and $s>0$,  suggesting that the Universe’s expansion was previously driven by a dark energy component with $-1<\omega<-0.33$. Moving toward the future, the trajectory slightly diverges from $\Lambda$CDM but remains close, implying that the model predicts a smooth evolution of dark energy without drastic deviations. The current values obtained from our model are: $r_{0}=0.5956$ and $s_{0}=0.1622$, which indicate a mild deviation from $\Lambda$CDM and support a slowly evolving dark energy component \cite{Singh23}. This trajectory validates the physical consistency of our model. It reproduces an early quintessence-like phase and approaches the expected behavior of the standard cosmological model at present. The result confirms that our model provides a dynamically viable description of late-time cosmic evolution, consistent with both theoretical expectations and current observational bounds.
\section{Age of the Universe for $f(T,T_{G})$ framework}\label{sec9}
\hspace{0.5cm} Accurately determining the age of the Universe serves as a crucial consistency test for any cosmological model. In our framework, the age is computed using the integral expression:
\begin{equation}\label{48}
t_{0}-t=\int_{0}^{z}\frac{dz}{(1+z)H(z)},
\end{equation}
Here, $t_{0}$ indicates the current cosmic age and $H(z$ is the model-derived Hubble parameter. By integrating our model's Hubble parameter (equation (\ref{25})) into the time–redshift relation, we derive the dimensionless time evolution function $H_{0}(t_{0}-t)$, which traces the Universe’s expansion history as a function of redshift.
\begin{equation}\label{49}
H_{0}(t_{0}-t)=\int_{0}^{z}\frac{dz}{(1+z)\sqrt{\bigg(\alpha e^{\frac{z}{1+z}}+\beta\bigg)\big[(1+z)^{3}-1\big]+1}}.
\end{equation}
\begin{figure}[hbt!]
  \centering
  \includegraphics[scale=0.4]{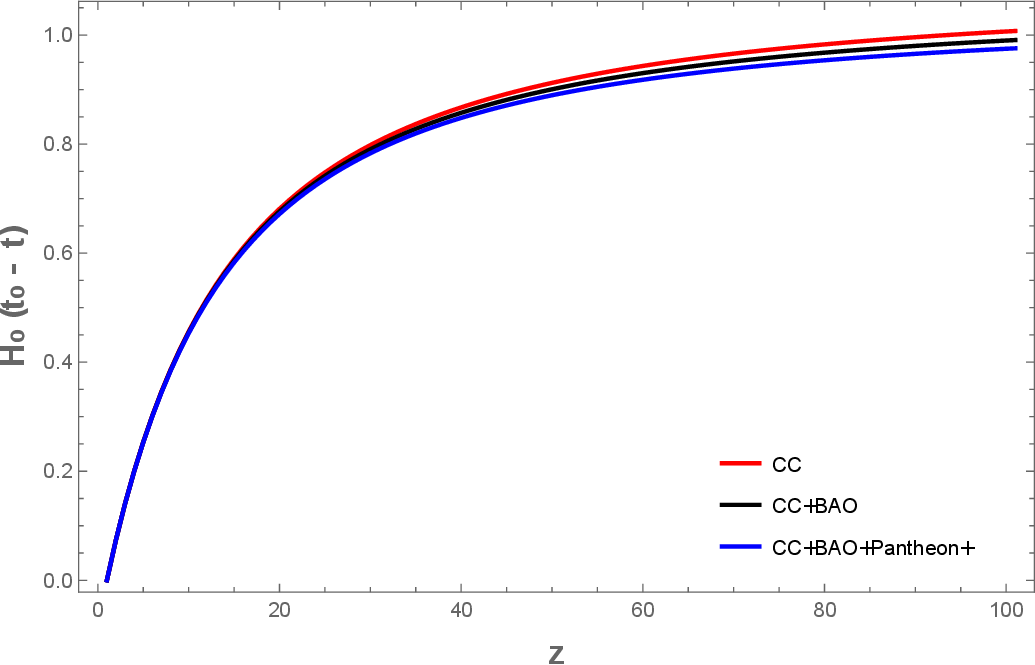}
  \caption{The variation of cosmic time with redshift for best fit values.}\label{fig:f9}
\end{figure}

By numerically evaluating this integral with our reconstructed $H(z)$, we obtain the dimensionless cosmic age in Hubble units, $H_{0}t_{0}$. The resulting values are: $H_{0}t_{0}=0.9987$ (CC), $0.9895$ (CC+BAO) and $0.9734$ (CC+BAO+Pantheon+). These correspond to physical age estimates of: $t_{0}\approx13.87$ Gyr (CC), $13.64$ Gyr (CC+BAO) and $13.28$ Gyr (CC+BAO+Pantheon+). Figure \ref{fig:f9} shows that the cosmic time evolution with redshift approaches a finite value at high redshifts, corresponding to the Universe's total age. Notably, all three values are in excellent agreement with recent astrophysical measurements—including those from Planck and globular cluster studies—which estimate the Universe’s age to lie between $13.5$ and $13.8$ Gyr. These results affirm the internal consistency of our $f(T,T_{G})$ model and reinforce its reliability in describing the late-time Universe. The predicted age not only falls within observational limits but also supports the model's compatibility with both cosmic expansion history and structure formation timescales \cite{Cruz20,IB20}.
\section{Conclusion}\label{sec10}
\hspace{0.5cm} In this study, we investigated the late-time cosmological implications of the modified teleparallel gravity model defined as $f(T,T_{G})=T+\gamma\sqrt{T_{G}}+\delta\sqrt{T}$. To constrain the model, we introduced an exponential form of the $Om(z)$ diagnostic, $Om(z)=\alpha e^{\frac{z}{1+z}}+\beta$ and derived the corresponding Hubble parameter $H(z)$. Using the MCMC method, we performed a statistical analysis based on three high-quality datasets: $31$ CC measurements, $26$ BAO points and $1701$ data points from the Pantheon+ Type Ia Supernovae compilation.

The MCMC analysis yielded the following parameter ranges: $H_{0}\in[68.46,77.387]$ km/s/Mpc, $\alpha\in[-0.232,-0.068]$ and $\beta\in[0.218,0.560]$. These values are in strong agreement with local Hubble constant measurements and suggest a mild deviation from the $\Lambda$CDM model. The negative values of $\alpha$ and positive values of $\beta$ point toward a slowly increasing $Om(z)$ function, indicating a phantom-like behavior of dark energy.

The model's predicted deceleration parameter $q(z)$ shows a smooth transition from early deceleration to late-time acceleration, with the transition redshift decreasing as more data are included. Current values of $q_{0}$ lie between $-0.340$ and $-0.323$, aligning with the observed accelerating expansion of the Universe.

Further, the analysis of energy density and pressure confirms physical viability: energy density remains positive throughout, while pressure becomes negative at low redshift—consistent with dark energy-driven acceleration. The evolution of the equation of state parameter $\omega(z)$ ransitions from stiff matter-like behavior at early times to quintessence-like behavior at present, with current values ranging from $\omega_{0}=-0.335$ to $-0.324$, avoiding the phantom regime.

The energy conditions show that NEC and DEC are respected, while SEC is violated—consistent with an accelerating phase. The statefinder diagnostic confirms that the model trajectory passes through the $\Lambda$CDM point in the $\{r,s\}$ plane, with current values $r_{0}=0.5956$ and $s_{0}=0.1622$, highlighting a smooth, observationally viable deviation.

Finally, the estimated age of the Universe derived from the model is consistent with astrophysical expectations: $t_{0}\approx13.87$ Gyr (CC), $13.64$ Gyr (CC+BAO) and $13.28$ Gyr (CC+BAO+Pantheon+). This confirms the robustness of the model and its capability to accurately describe the Universe’s expansion history.

In summary, our proposed exponential $Om(z)$ parametrization within the framework of $f(T,T_{G})$ gravity provides a dynamically consistent and observationally validated description of late-time cosmic acceleration. The assumed model exhibits stable and physically realistic behavior across all epochs, accurately tracing the Universe’s transition from deceleration to acceleration. Moreover, the framework offers flexibility for future extensions, including the exploration of inflationary dynamics, perturbation growth, or interactions with dark matter within the modified gravity setup.
 
 \end{document}